\documentclass[preprint]{sigplanconf}

\usepackage{amsmath}
\usepackage{graphicx}
\usepackage{float}
\usepackage{textcomp}
\usepackage[pdfborder={0 0 0}]{hyperref}

\begin{document}

\conferenceinfo{International Symposium on Memory Management 2011}{June 4th-5th 2011, San Jose, U.S.A.} 
\copyrightyear{2010-2011} 
\copyrightdata{Niall Douglas} 

\titlebanner{Draft 4}        % These are ignored unless
\preprintfooter{Draft 4}   % 'preprint' option specified.

\title{User Mode Memory Page Allocation}
\subtitle{A Silver Bullet For Memory Allocation?}

\authorinfo{Mr. Niall Douglas MBS MA BSc}
           {ned Productions IT Consulting}
           {\url{http://www.nedproductions.biz/}}
\hypersetup{pdftitle={User Mode Memory Page Allocation: A Silver Bullet For Memory Allocation?}, pdfauthor={Niall Douglas}, pdfsubject={International Symposium on Memory Management 2011}, pdfkeywords={MMU, mmap, sbrk, malloc, realloc, free, O(1), faster array extension, bare metal}}

\maketitle

\begin{abstract}
The literature of the past decade has discussed a number of contrasting ways in which to improve general allocation performance. The Lea allocator, dlmalloc \cite{lea2000memory}, aims for a reusable simplicity of implementation in order to improve third-party customisability and performance, whereas other allocators have a much more complex implementation which makes use of per-processor heaps, lock-free, cache-line locality and transactional techniques \cite{berger2000hoard, michael2004scalable, hudson2006mcrt}. Many still believe strongly in the use of custom application-specific allocators despite that research has shown many of these implementations to be sub-par \cite{berger2002reconsidering}, whereas others believe that the enhanced type information and metadata available to source compilers allow superior allocators to be implemented at the compilation stage \cite{berger2001composing, udayakumaran2003compiler, boostpool2010}.

More recently there appears to be an increasingly more holistic view of how the allocator's implementation strategy non-linearly affects whole application performance via cache set associativity effects and cache line localisation effects, with a very wide range of approaches and techniques \cite{hudson2006mcrt, dice2010simplifying, Dice2010}. This paper extends this more holistic view by examining how the design of the paged virtual memory system interacts with overall application performance from the point of view of how the growth in memory capacity will continue to exponentially outstrip the growth in memory speed, and the consequent memory allocation overheads that will be thus increasingly introduced into future application memory allocation behaviour.

The paper then proposes a novel solution: the elimination of paged virtual memory and partial outsourcing of memory page allocation and manipulation from the operating system kernel into the individual process' user space -- \emph{a user mode page allocator} -- which allows an application to have direct, bare metal access to the page mappings used by the hardware Memory Management Unit (MMU) for its part of the overall address space. A user mode page allocator based emulation of the \texttt{mmap()} abstraction layer of dlmalloc is then benchmarked against the traditional kernel mode implemented \texttt{mmap()} in a series of synthetic Monte-Carlo and real world application settings.

Despite the highly inefficient implementation, scale invariant performance up to 1Mb block sizes was found in the user mode page allocator, which translates into up to a 10x performance increase for manipulation of blocks sized 0-8Mb and up to a 4.5x performance increase for block resizing. A surprising 2x performance increase for very small allocations when running under non-paged memory was also found, and it is speculated that this could be due to cache pollution effects introduced by page fault based lazy page allocation.

In real world application testing where dlmalloc was binary patched in to replace the system allocator in otherwise unmodified application binaries (i.e. a worst case scenario), every test scenario, bar one, saw a performance improvement with the improvement ranging between -4.05\% and +5.68\% with a mean of +1.88\% and a median of +1.20\%.

Given the superb synthetic and positive real world results from the profiling conducted, this paper proposes that with proper operating system and API support one could gain a further order higher performance again while keeping allocator performance invariant to the amount of memory being allocated or freed i.e. a \emph{\textbf{100x}} performance improvement or more in some common use cases. It is rare that through a simple and easy to implement API and operating system structure change one can gain a Silver Bullet with the potential for a second one.
\end{abstract}




\section{Introduction}

\subsection{Historical trends in growths of capacities and speeds}

\begin{figure}[h]
  \centering
    \includegraphics[width=0.5\textwidth]{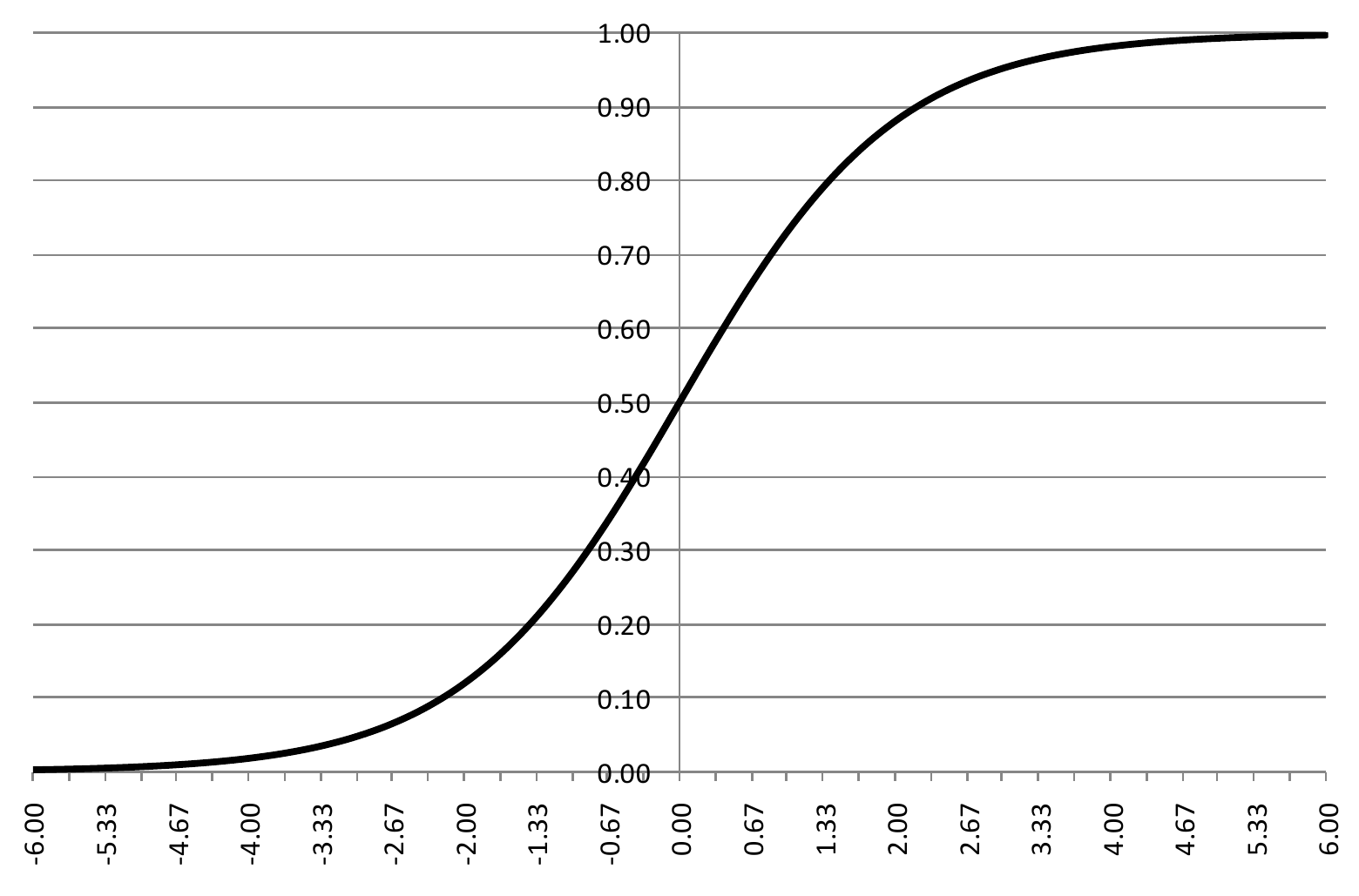}
  \caption{A plot of the standard logistic growth curve $\frac{1}{1+e^{-t}}$.}
  \label{FigLogisticGrowthCurve}
\end{figure}

As described by J.C. Fisher in his seminal 1972 paper \cite{fisher1972simple} and reconfirmed in much research since \cite{meyer1994bi, meyer1999carrying}, technological growth is never better than logistic in nature. Logistic growth, often called an ``S-curve", occurs when growth is introduced by a feedback loop whereby the thing grown is used to grow more of that thing -- which is implicitly a given when new computer technology is used to make available the next generation of computer technology. As with any feedback loop driven growth, hard ``carrying capacity" limits such as the fixed physical size of the silicon atom eventually emerge which decrease the initial exponential rate of growth down to linear, then to exponentially declining growth i.e. the rate of growth of growth -- the second derivative -- goes negative long before the rate of growth itself does. Because feedback driven growth tends to follow a logistic curve, as Fisher showed one cannot predict when exponential growth will end until the first derivative goes negative for an extended period of time -- and only now can one predict with increasing probability when overall growth will cease. This pattern of growth is summarised in Figure \ref{FigLogisticGrowthCurve} where 0.0 along the x-axis shows the point of inflection where growth turns negative.

\begin{figure}[h]
  \centering
    \includegraphics[width=0.5\textwidth]{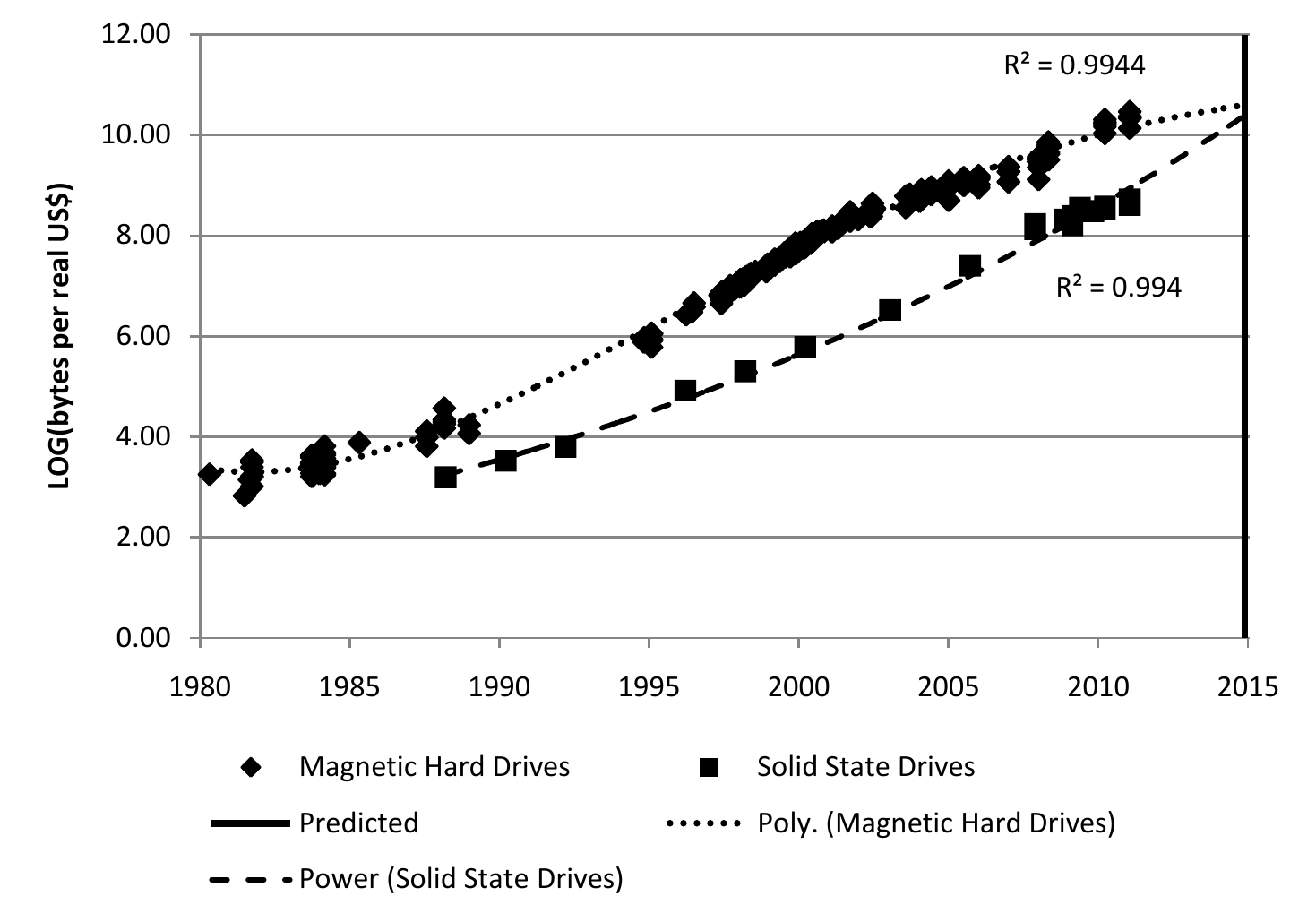}
  \caption{A log plot of bytes available per inflation adjusted US\$ from 1980 to 2010 for conventional magnetic hard drives and flash based solid state disk drives. Magnetic hard drives are clearly coming to the end of their logistic growth curve trajectory, whereas flash drives are still to date undergoing the exponential growth part of their logistic growth curve. Sources: \cite{winch2010} \cite{wiki2010b} \cite{storsearch2010}.}
  \label{FigSSDsVsHardDrives}
\end{figure}

The history of growth in computer technology has been nothing short of remarkable in the last five decades, but one must always remember that nothing can grow exponentially forever \cite{meyer1999primer}. In the growth of storage capacity and transistor density we are still in the part of the logistic growth curve before the mid-turning-point where growth appears exponential and no one can yet say when the point of inflection (when the rate of growth of growth i.e. the second derivative turns negative) will come. In comparison, magnetic hard drive storage is not just past its point of inflection (which was in 1997), but it is rapidly approaching zero growth which Figure \ref{FigSSDsVsHardDrives} makes all too obvious. All other things being equal and if present growth trends continue, flash based non-volatile storage ought to overtake magnetic based non-volatile storage some time in 2013 plus or minus eighteen months.

The implications of this change are \textbf{profound} for all users of computing technology, but especially for those responsible for the implementations of system memory management. As Denning (1996) \cite{denning1996} recounts most vividly, from the 1950s onward the present system of memory management was designed on the basis that non-volatile storage was large but with a high access latency, whereas volatile storage (RAM) was small but with a low access latency. The problem was always one of how to make best use of the limited low-latency volatile storage in making use of as much of the high-latency non-volatile storage as possible -- thus the modern automatic paged virtual storage allocation system (hereafter called ``paged virtual memory") with the concept of a ``working set" was born, and which was described by Denning himself in his seminal 1970 paper \cite{denning1970virtual}.

As a much simplified explanation of the essentials of Denning's paper, what happens is that requests for memory from the operating system kernel allocate that memory on the non-volatile storage (typically in the paging, or `swap' file used to temporarily store presently least used volatile memory pages) and the kernel returns a region of virtual address space of the requested size. Note that no physical RAM is allocated at this stage. When the process first accesses a RAM page (the granularity of the MMU's page mapping, typically no smaller than 4Kb on modern systems), the CPU triggers a \emph{page fault} as the memory does not exist at that address. The operating system kernel then takes a free physical RAM page and places it at the site of the fault before restoring normal operation. If a previously accessed RAM page is not used for a long time and if new free RAM pages are needed, then the contents of that page are written off into the paging file and its physical RAM page used for something more frequently used instead. This allows only those pages which are most frequently used to consume physical RAM, thus reducing the true RAM consumption of the process as far as possible while still allowing the process to conceptually use much more memory than is physically present in the computer. Almost all computer systems with any complexity use such a system of paged virtual memory, or a simplification thereof.

\subsection{The first problem with page file backed memory}

\begin{figure}[h]
  \centering
    \includegraphics[width=0.5\textwidth]{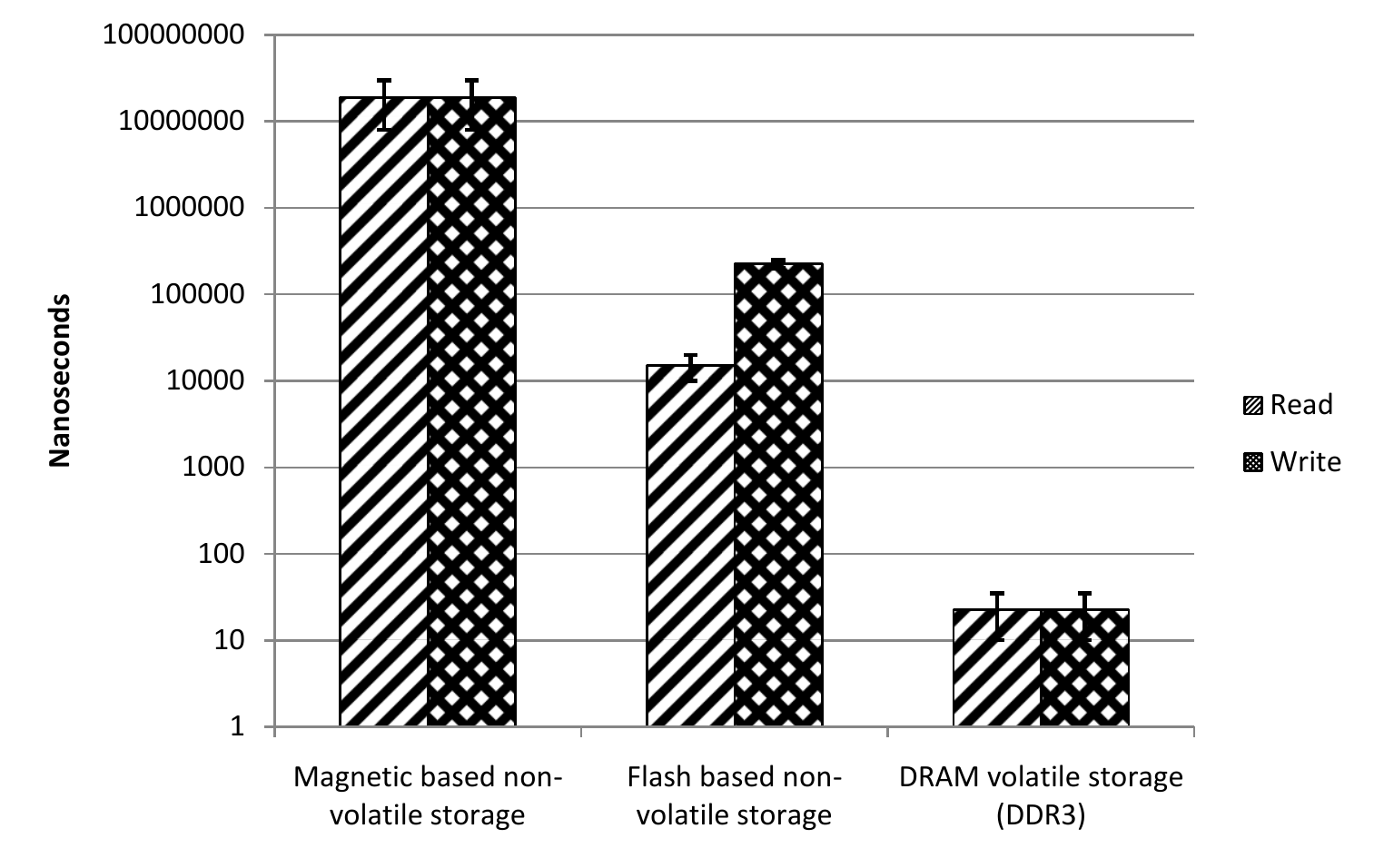}
  \caption{A log plot of how much time is required to randomly access selected computer data storage media. Sources: \cite{tomshardware2010, desnoyers2010empirical}.}
  \label{FigStorageLatencies}
\end{figure}

The first problem given the trends outlined in Figure \ref{FigSSDsVsHardDrives} with paged virtual memory has already been shown: the fact that high latency magnetic non-volatile storage is due to become replaced with flash based storage. Unlike magnetic storage whose average random access latency may vary between 8-30ms \cite{tomshardware2010} and which has a variance strongly dependent on the distance between the most recently accessed location and the next location, flash based storage has a flat and uniform random access latency just like RAM. Where DDR3 RAM may have a 10-35ns read/write latency, current flash storage has a read latency of 10-20\textmu s, a write latency of 200-250\textmu s (and an erase latency of 1.5-2ms, but this is usually hidden by the management controller) \cite{desnoyers2010empirical} which is only 667x slower than reading and 10,000x slower than writing RAM respectively. What this means is that the relative overhead of the page file backed memory system implementation on flash based storage becomes \emph{much larger} -- several thousand fold larger -- against overall performance than when based on magnetic storage based swap files.

Furthermore flash based storage has a limited write cycle lifetime which makes it inherently unsuitable for use as temporary volatile storage. Thankfully due to the exponential growth in RAM capacity per currency unit, RAM capacity is in abundance on modern systems -- the typical new home computer today in 2010 comes with far more RAM than the typical home user will ever use, and RAM capacity constraints are only really felt in server virtualisation, web service, supercomputing and other ``big iron" applications.

\subsection{The second problem with page file backed memory}
\begin{figure}[h]
  \centering
    \includegraphics[width=0.5\textwidth]{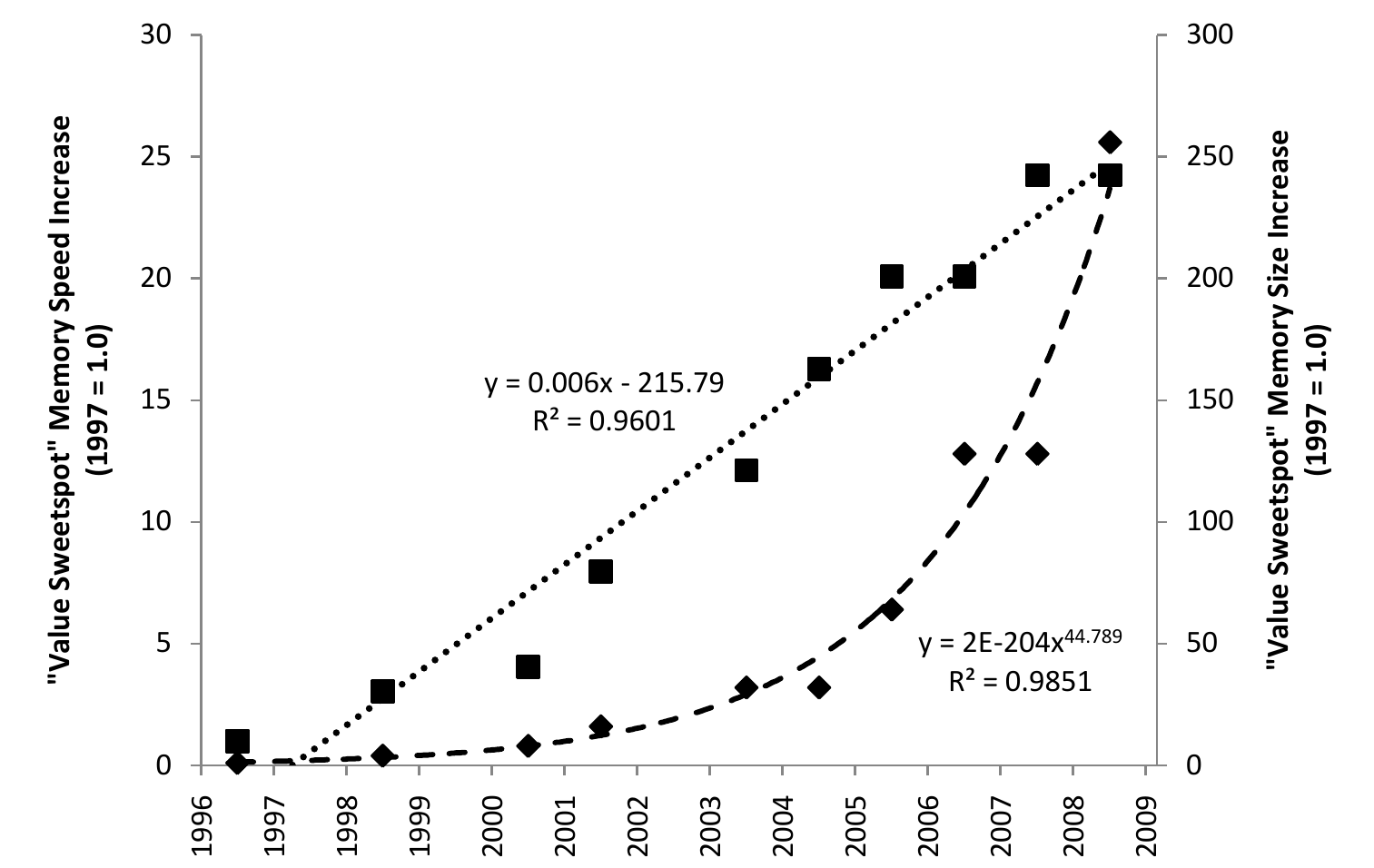}
  \caption{A plot of the relative growths since 1997 of random access memory (RAM) speeds and sizes for the best value (in terms of Mb/US\$) memory sticks then available on the US consumer market from 1997 - 2009, with speed depicted by squares on the left hand axis and with size depicted by diamonds on the right hand axis. The dotted line shows the best fit regression for speed which is linear, and the dashed line shows the best fit for size which is a power regression. Note how that during this period memory capacity outgrew memory speed by a factor of ten. Sources: \cite{jcmit2010} \cite{wiki2010}.}
  \label{FigMemorySizeVsSpeed}
\end{figure}

This introduces the second problem with paged virtual memory: it was designed under the assumption that low latency non-volatile storage capacity is scarce and whose usage must be absolutely maximised, and that is increasingly not true. As Figure \ref{FigMemorySizeVsSpeed} shows, growth in capacity for the ``value sweetspot" in RAM (i.e. those RAM sticks with the most memory for the least price at that time) has outstripped growth in the access speed of the same RAM by a factor of ten in the period 1997-2009, so while we have witnessed an impressive 25x growth in RAM speed we have also witnessed a \textbf{250x} growth in RAM capacity during the same time period. Moreover, growth in capacity is exponential versus a linear growth in access speed, so the differential is only going to dramatically increase still further in the next decade. Put another way, assuming that trends continue and all other things being equal, if it takes a 2009 computer 160ms to access all of its memory at least once, it will take a 2021 computer \textbf{5,070 years} to do the same thing\footnote{No one is claiming that this will actually be the case as the exponential phase of logistic growth would surely have ended. However, even if capacity growth goes linear sometime soon, one would still see the growth in capacity outpace the growth in access speed by anywhere between tenfold and one trillion ($10^{12}$) times (for reference, the latter figure was calculated as follows: (i) double period of exponential growth since 1997 $log_{44.8}(250) = 1.131$, $(2 \times 1.131)^{44.8} = 7.657 \times 10^{15}$ (ii) remove exponential growth 1997-2009 $\frac{7.657 \times 10^{15}}{250}$ = $3 \times 10^{13}$ (iii) divide exponential capacity growth by linear access speed growth 2009-2021 $\frac{3 \times 10^{13}}{25}$ = $1.2 \times 10^{12}$).}.

From the growth trends evident in these time series data, one can make the following observations:
\begin{enumerate}
\item Recently what is much more scarce than capacity is \emph{access speed} -- as anyone who has recently tried a full format of a large hard drive can tell you, the time to write once to the entire capacity is also exponentially growing because the speed of access is only growing linearly.
\item If the growth in RAM capacity over its access speed continues, we are increasingly going to see applications constrained not by insufficient storage, but by insufficiently fast access \emph{to} storage.
\item It is reasonable to expect that programmers will continue to expend RAM capacity usage for improvements in other factors such as application performance, development time and user experience. In other words, the ``working set" of applications is likely to also grow as exponentially as it can (though one would struggle to see how word processing and other common tasks could ever make good use of huge capacities). This implies that the overheads spent managing capacity will also grow exponentially.
\end{enumerate}

Considering all of the above, one ought to therefore conclude that, algorithmically speaking, \textbf{capacity ought to be expended wherever latency can be reduced}.

\subsection{Paged virtual memory overheads}

\begin{figure}[h]
  \centering
    \includegraphics[width=0.5\textwidth]{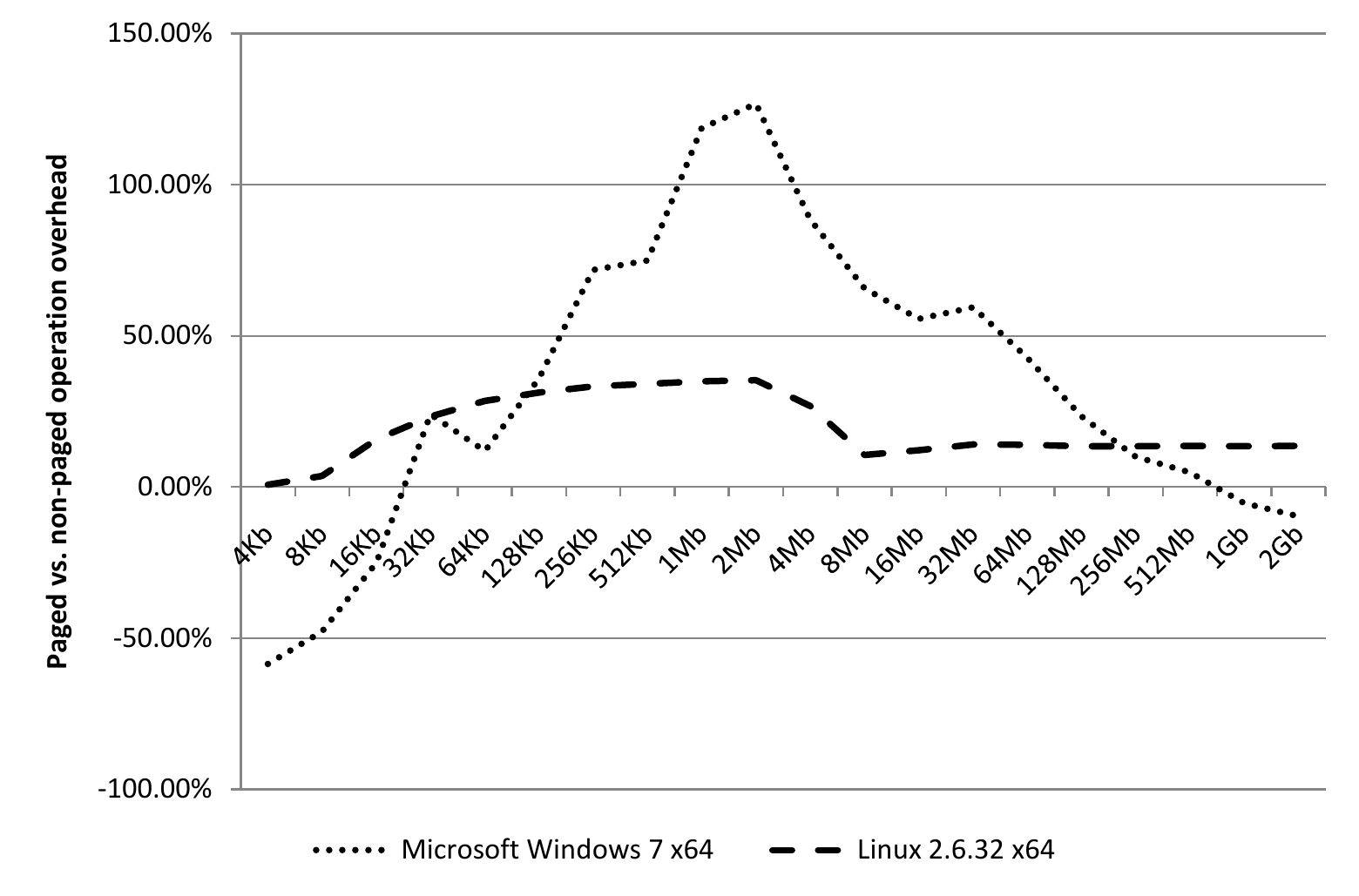}
  \caption{A log plot of how much overhead paged virtual memory allocation introduces over non-paged memory allocation according to block size.}
  \label{FigPagedMemoryOverheads}
\end{figure}

Figure \ref{FigPagedMemoryOverheads} shows how much overhead is introduced in a best case scenario by fault driven page allocation versus non-paged allocation for Microsoft Windows 7 x64 and Linux 2.6.32 x64 running on a 2.67Ghz Intel Core 2 Quad processor\footnote{For reference, this system has 32Kb of L1 data cache with 16 entry L1 TLB, 4Mb of L2 cache with 256 entry L2 TLB \cite{intelq6000} and 6Gb of DDR2 RAM with 12.8Gb/sec maximum bandwidth. The L1 TLB can hold page mappings for 64Kb of memory, whereas the L2 TLB can hold mappings for 1Mb of memory \cite{intelq6000}.}. The test allocates a block of a given size and writes a single byte in each page constituting that block, then frees the block. On Windows the Address Windowing Extension (AWE) functions \texttt{AllocateUserPhysicalPages()} et al. were used to allocate non-paged memory versus the paged memory returned by \texttt{VirtualAlloc()}, whereas on Linux the special flag \texttt{MAP\_POPULATE} was used to ask the kernel to prefault all pages before returning the newly allocated memory from \texttt{mmap()}. As the API used to perform the allocation and free is completely different on Windows, one would expect partially incommensurate results.

\begin{figure}[h]
  \centering
    \includegraphics[width=0.5\textwidth]{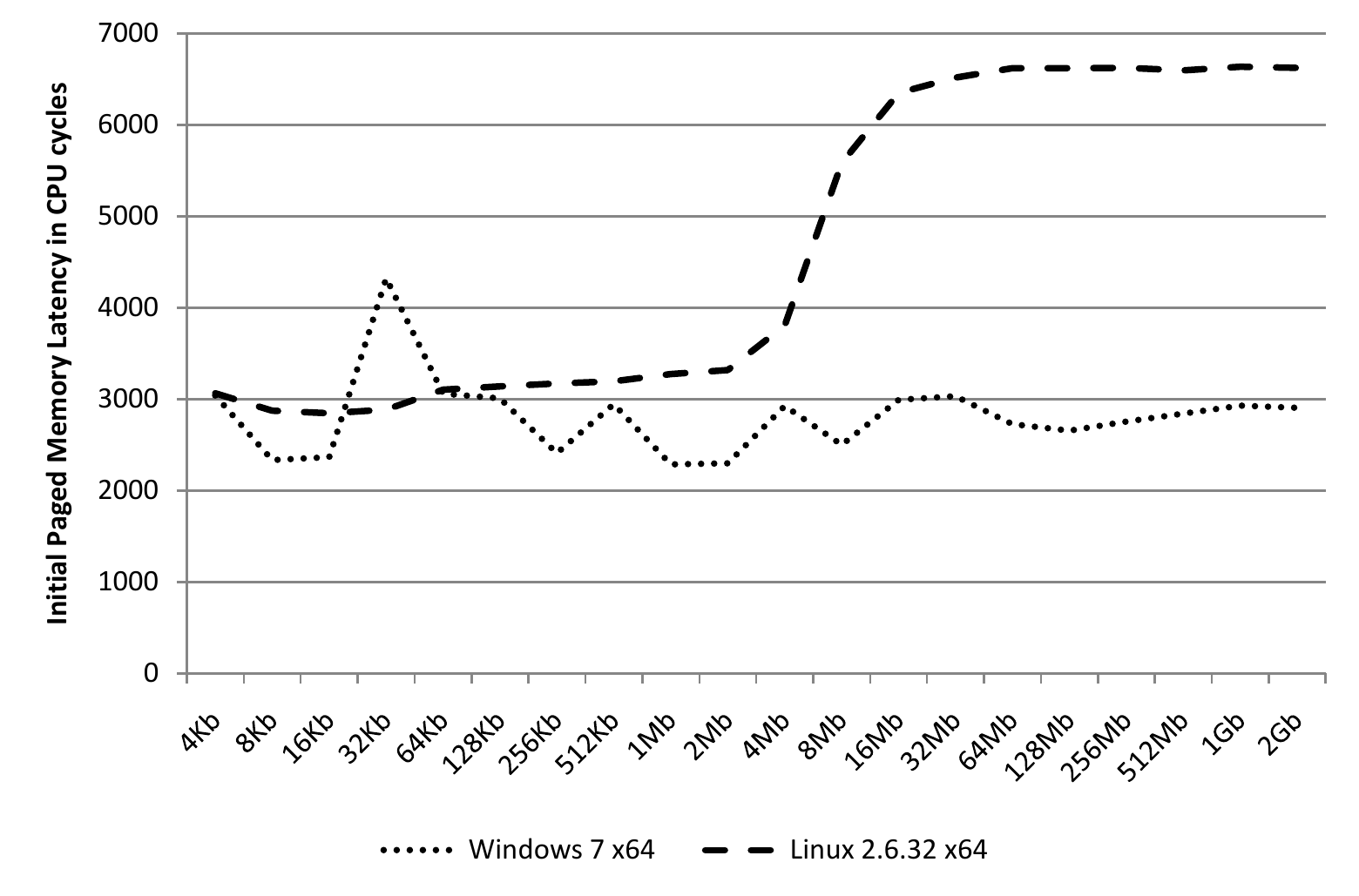}
  \caption{A log plot of how many CPU cycles is consumed per page by the kernel page fault handler when allocating a run of pages according to block size.}
  \label{FigKernelPageFaultHandlerPerformance}
\end{figure}

As one can see, the overhead introduced by fault driven page allocation is substantial with the overhead reaching 100\% on Windows and 36\% on Linux. The overhead rises linearly with the number of pages involved up until approximately 1-2Mb after which it drops dramatically. This makes sense: for each page fault the kernel must perform a series of lookups in order to figure what page should be placed where, so the overhead from the kernel per page fault as shown in Figure \ref{FigKernelPageFaultHandlerPerformance} ought to be approximately constant at 2800 cycles per page for Microsoft Windows 7. There is something wrong with the Linux kernel page fault handler here: it costs 3100 cycles per page up to 2Mb which seems reasonable, however after that it rapidly becomes 6500 cycles per page which suggests a TLB entry size dependency\footnote{For reference, when the CPU accesses a page not in its Translation Lookaside Buffer (TLB) it costs a page table traversal and TLB entry load which on this test machine is approximately 15 cycles when the page is on the same page table sheet, and no more than 230 cycles if it is in a distant sheet.} or a lack of easily available free pages and which is made clear in Table \ref{TableMemoryLatencies}. However that isn't the whole story -- obviously enough, each time the kernel executes the page fault handler it must traverse several thousand cycles worth of code and data, thus kicking whatever the application is currently doing out of the CPU's instruction and data caches and therefore necessitating a reload of those caches when execution returns.

In other words, fault driven page allocation introduces a certain amount of ongoing CPU cache pollution and therefore \emph{raises the latency of first access to a memory page}.

\begin{table}[hb]
\caption{Selected page fault allocation latencies for a run of pages on Microsoft Windows and Linux.}
\begin{center}
\begin{tabular}{ r | r r | r r }
& \multicolumn{2}{|c|}{Microsoft Windows} & \multicolumn{2}{|c}{Linux} \\
& \small{Paged} & \small{Non-paged} & \small{Paged} & \small{Non-paged} \\
Size & \small{cycles/page} & \small{cycles/page} & \small{cycles/page} & \small{cycles/page} \\
\hline
16Kb & 2367 & 14.51 & 2847 & 15.83 \\
1Mb & 2286 & 81.37 & 3275 & 14.53 \\
16Mb & 2994 & 216.2 & \textbf{6353} & 113.4 \\
512Mb & 2841 & 229.9 & \textbf{6597} & 115.9 \\
\end{tabular}
\end{center}
\label{TableMemoryLatencies}
\end{table}

The point being made here is that paged virtual memory introduces a lot of additional and often hard to predict application execution latency through the entire application as a whole when that application makes frequent changes to its virtual address space layout. The overheads shown in Figures \ref{FigPagedMemoryOverheads} and \ref{FigKernelPageFaultHandlerPerformance} represent a best-case scenario where there isn't a lot of code and data being traversed by the application -- in real world code, the additional overhead introduced by pollution of CPU caches by the page fault handler can become sufficiently pathological that some applications deliberately prefault newly allocated memory before usage.

The typical rejoinder at this point is that MMUs do allow much larger page sizes (2Mb on x64) to be concurrently used with smaller ones, so an improved fault driven page allocator could map 2Mb sized pages if its heuristics suggest that to be appropriate. However as anyone who has tried to implement this knows, this is very tricky to get right \cite{navarro2002practical}. Moreover, one can only detect when an individual page has been modified (becomes `dirty') and so therefore the use of large pages can generate very significant extra i/o when writing out dirty pages when pages are backed by a non-volatile store.

\subsection{Code complexity}
Since the 1970s, application and kernel programmers have assumed that the larger the amount of memory one is working with, the longer it will take simply because it has always been that way historically -- and after all, logically speaking, it must take ten times as many opcodes executed in order to access ten times as much memory by definition. Therefore, programmers will tend to be more likely to instinctively dedicate significant additional code complexity to reducing memory usage as the amount of memory usage grows due to the standard space/time trade-off assumptions made by all computer programmers since the very earliest days (e.g. \cite{bloom1970space}). In other words, if memory usage looks ``too big" relative to the task being performed, the programmer will tend to increase the complexity of code in order to reduce the usage of memory -- a `balancing' of resource usage as it were.

However, the standard assumptions break down under certain circumstances. In the past, the speed of the CPU was not much different from that of memory, so memory access costs were strongly linked to the number of opcodes required to perform the access. As CPUs have become ever more faster than main memory, such is today's disjunct between CPU speed and memory speed that (as indeed we shall shortly see) \textbf{scale invariant memory allocation has become possible}. In other words, today's CPUs run so much faster than memory that they can allocate gigabytes of memory in the same time as a few dozen kilobytes \emph{if the implementation is written to avoid CPU instruction dependencies upon memory}.

The problem with the assumption that larger memory usage takes longer is that it introduces significant extra complexity into both the kernel and the medium to large sized application which \textbf{most programmers are so used to that they don't even think about it}. Garbage collecting, reference counting, least frequently used caching -- all are well known as object lifetime management techniques in the application space, while in the kernel space much effort is dedicated to trying to extract a ``one size fits all" set of statistical assumptions about typical memory usage patterns of the average application and writing a paging virtual memory system to attempt an optimal behaviour based on predicted working set.

Just to show the ludicrousness of the situation, consider the following: most applications use a C malloc API (whose design is almost unchanged since the 1970s) which takes no account of virtual memory \emph{at all} -- each allocates memory as though it is the sole application in the system. The kernel then \emph{overrides} the requested memory usage of each process on the basis of a set of statistical assumptions about which pages are actually used by \emph{faking} an environment for the process where it is alone in the system and can have as much memory as it likes. Meanwhile you have programmers finding pathological performance in some corner cases in this environment, and writing code which is specifically designed to fool the kernel into believing that memory is being used when it is not, despite that such prefaulting hammers the CPU caches and is fundamentally a waste of CPU time and resources.

\section{User mode page allocation as a potential solution}
The solution proposed by this paper to the problems outlined above is simple:
\begin{enumerate}
\item Except for memory mapped files, do away with kernel paged virtual memory altogether\footnote{Ideally, even memory mapped files ought to be a user mode only implementation. Strictly speaking, even shared memory mapped regions used for inter-process communications only need the kernel for construction and destruction of the shared memory region only -- thereafter a user mode filing system implementation using the kernel page fault handler hooks could do the rest.}! In other words, when you map a RAM page to some virtual address, you get an actual, real physical RAM page at that address. This should allow a significant -- possibly a very significant on some operating systems -- reduction in the complexity of virtual memory management implementation in the kernel. Less kernel code complexity means fewer bugs, better security and more scalability \cite{curtis1979third, kearney1986software, khoshgoftaar1990predicting, banker1998software}.
\item Provide a kernel syscall which returns an arbitrary number of physical memory pages with an arbitrary set of sizes as their raw physical page frame identifiers as used directly by the hardware MMU. The same call can both free and request physical pages simultaneously, so a set of small pages could be exchanged for large pages and so on.
\item If the machine has sufficiently able MMU hardware e.g. NPT/EPT on x64, expose to user mode code direct write access to the page tables for each process such that the process may arbitrarily remap pages within its own address space to page frames it is permitted to access\footnote{In other words, one effectively virtualises the MMU for all applications as a hardware-based form of page mapping security enforcement.}. If without a sufficiently able MMU, a kernel call may be necessary to validate the page frames before permitting the operation.
\item Allow user mode code to issue TLB flushes if necessary on that CPU architecture.
\item Provide two asynchronous kernel upcalls such that (i) the kernel can ask the process to release unneeded pages according to a specified ``severity" and (ii) that the kernel can notify a handler in the process of CPU access to a watched set of pages.
\item Rewrite the family of APIs \texttt{VirtualAlloc()} on Windows and \texttt{mmap(MAP\_ANONYMOUS)} on POSIX to be simple user mode wrappers around the above functionality i.e. get requisite number of physical pages and map them at some free virtual address. In other words, virtual address space management for a process becomes exclusively the process' affair. Add new APIs permitting asynchronous batch modifications (more about this shortly).
\item For backwards compatibility, and for the use of applications which may still use much more memory than is physically present in the machine, the process' memory allocator ought to provide a page file backed memory store in user mode implemented using the facilities above. This would be easy to implement as one can still memory map files and/or use the installable page fault kernel handlers mentioned earlier, and whose implementation could be made much more flexible and tunable than at present especially as it opens the possibility of competing third-party user mode implementations.
\end{enumerate}

In other words, direct access to the page mapping tables and hardware is handed over to the application. Because the application can now keep a lookaside cache of unused memory pages and it can arbitrarily relocate pages from anywhere to anywhere within its virtual address space, it can make the following primary optimisations in addition to removing the CPU cache pollution and latencies introduced by page faulted allocation:
\begin{enumerate}
\item An allocated memory block can be very quickly extended or shrunk without having to copy memory -- a feature which is \emph{very} useful for the common operation of extending large arrays and which is also provided by the proprietary \texttt{mremap()} function under Linux. \citet{kimpe2006} researched the performance benefits of a vector class based upon this feature and found a 50-200\% memory usage overhead when using a traditional vector class over a MMU-aware vector class as well as extension time complexity becoming dependent on the elements being added rather than the size of the existing vector. While the test employed was synthetic, a 50\% improvement in execution time was also observed thanks to being able to avoid memory copying.
\item Two existing memory blocks can have their contents swapped without having to copy memory. Generally programmers work around the lack of this capability by swapping pointers to memory rather than the memory itself, but one can see potential in having such a facility.
\item The application has \emph{far} better knowledge than any kernel could ever have about what regions of memory ought to be mapped within itself using larger pages. In combination with the lack of page file backed storage, user mode page allocation could enable significant TLB optimisations in lots of ways that traditional paged virtual memory never could.
\item Lastly, and most importantly of all, \emph{one no longer needs to clear contents when allocating new memory pages} if there are free pages in the lookaside cache. To explain why this is important, when a process returns memory pages to the kernel it could be that these pages may contain information which may compromise the security of that process and so the kernel must zero (clear) the contents of those pages before giving them to a new process. Most operating system kernels try to delay page clearing until when the CPU is idle which is fine when there is plenty of free RAM and the CPU has regular idle periods, however it does introduce significant physical memory page fragmentation which is unhelpful when trying to implement large page support, and besides in the end the CPU and memory system are being tied up in a lot of unnecessary work which comes with costs as non-obvious as increased electrical power consumption. User mode page allocation lets the memory allocator of the process behave far more intelligently and only have pages cleared when they need to be cleared, thus increasing whole application performance and performance-per-watt efficiency \textbf{especially} when the CPU is constantly maxed out and page clears must be done on the spot with all the obvious consequences for cache pollution.
\end{enumerate}

\section{Raw kernel and user page allocator performance}
In the end though, theory is all well and good, but will it actually work and will it behave the way in which theory suggests? To discover the answer to this, a prototype user mode page allocator was developed which abuses the Address Windowing Extensions (AWE) API of Microsoft Windows mentioned earlier in order to effectively provide direct access to the hardware MMU. The use of the verb `abuses' is the proper one: the AWE functions are intended for 32-bit applications to make use of more than 4Gb of RAM, and they were never intended to be used by 64 bit applications in arbitrarily remapping pages around the virtual address space. Hence due API workarounds the prototype user mode page allocator runs at least 10x slower than it ought to were the API more suitable\footnote{The reference to API workarounds making the prototype user mode page allocator 10x slower refers specifically to how due to API restrictions, page table modifications must be separated into two buffers one of which clears existing entries, and the second maps new entries. In addition to these unnecessary memory copies two separate calls must be made to \texttt{MapUserPhysicalPages()} to clear and \texttt{MapUserPhysicalPagesScatter()} to set despite that this ought to be a single call. In testing, it was found that this buffer separation and making of two API calls is approximately 8x-12x slower than a single direct \texttt{MapUserPhysicalPagesScatter()} call.}, and probably more like 40x slower when compared to a memory copy directly into the MMU page tables -- however, its dependency or lack thereof on allocation size in real world applications should remain clear.

As suggested above, the user mode page allocator provides five APIs for external use with very similar functionality to existing POSIX and Windows APIs: \texttt{userpage\_malloc()}, \texttt{userpage\_free{}}, \texttt{userpage\_realloc()}, \texttt{userpage\_commit()} and \texttt{userpage\_release()}. This was done to ease the writing of kernel API emulations for the binary patcher, described later in this paper.

\begin{figure}[h]
  \centering
    \includegraphics[width=0.5\textwidth]{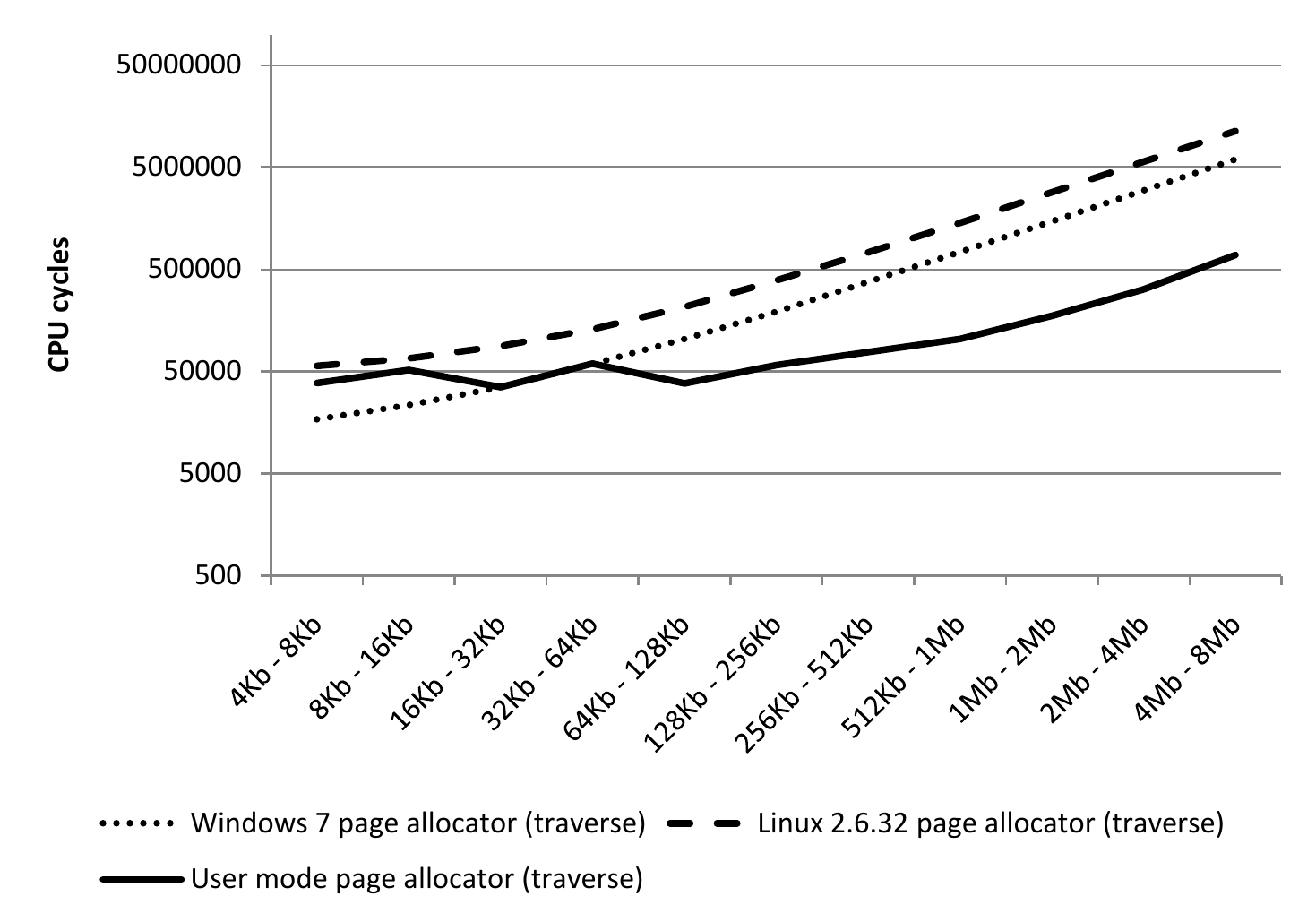}
  \caption{A log-log plot of how the system and user mode page allocators scale to block size.}
  \label{FigSystemVsUserMode}
\end{figure}

Figure \ref{FigSystemVsUserMode} shows on a log-log plot how the Windows 7 x64 kernel page allocator, Linux 2.6.32 x64 kernel page allocator and user mode page allocators scale to allocation block size. This test chooses a random number between 4Kb and 8Mb and then proceeds to request an allocation for that size and write a single byte into each page of it. Each allocated block is stored in a 512 member ring buffer and when the buffer is full the oldest item is deallocated. The times for allocation, traversal and freeing are then added to the two powers bin for that block size and the overall average value is output at the end of the test. Before being graphed for presentation here, the results are adjusted for a null loop in order to remove all overheads not associated with allocator usage, and where an item says `notraverse' it means that it has been adjusted for the time it takes for the CPU to write a single byte into each page in the allocation i.e. notraverse results try to remove as much of the effects of memory traversal as possible. Note that all following results for the user mode page allocator are for a best case scenario where its lookaside page cache can supply all requests i.e. the test preloads the free page cache before testing.

As would be expected, Figure \ref{FigSystemVsUserMode} shows that page allocation performance for the kernel page allocators is very strongly linearly correlated with the amount of memory being allocated and freed with apparently no use of pre-zeroed pages on either Windows or Linux\footnote{According to a year 2000 email thread \cite{eggert2000} on the FreeBSD mailing list, idle loop pre-zeroing of pages was benchmarked in Linux x86 and found to provide no overall performance benefit, so it was removed. Apparently the PowerPC kernel build does perform idle loop pre-zeroing however.}. Meanwhile, the user mode page allocator displays an apparent scale invariance up to 1Mb before also becoming linearly correlated with block size, with performance being lower up to 128Kb block sizes due to inefficiency of implementation.

\begin{figure}[h]
  \centering
    \includegraphics[width=0.5\textwidth]{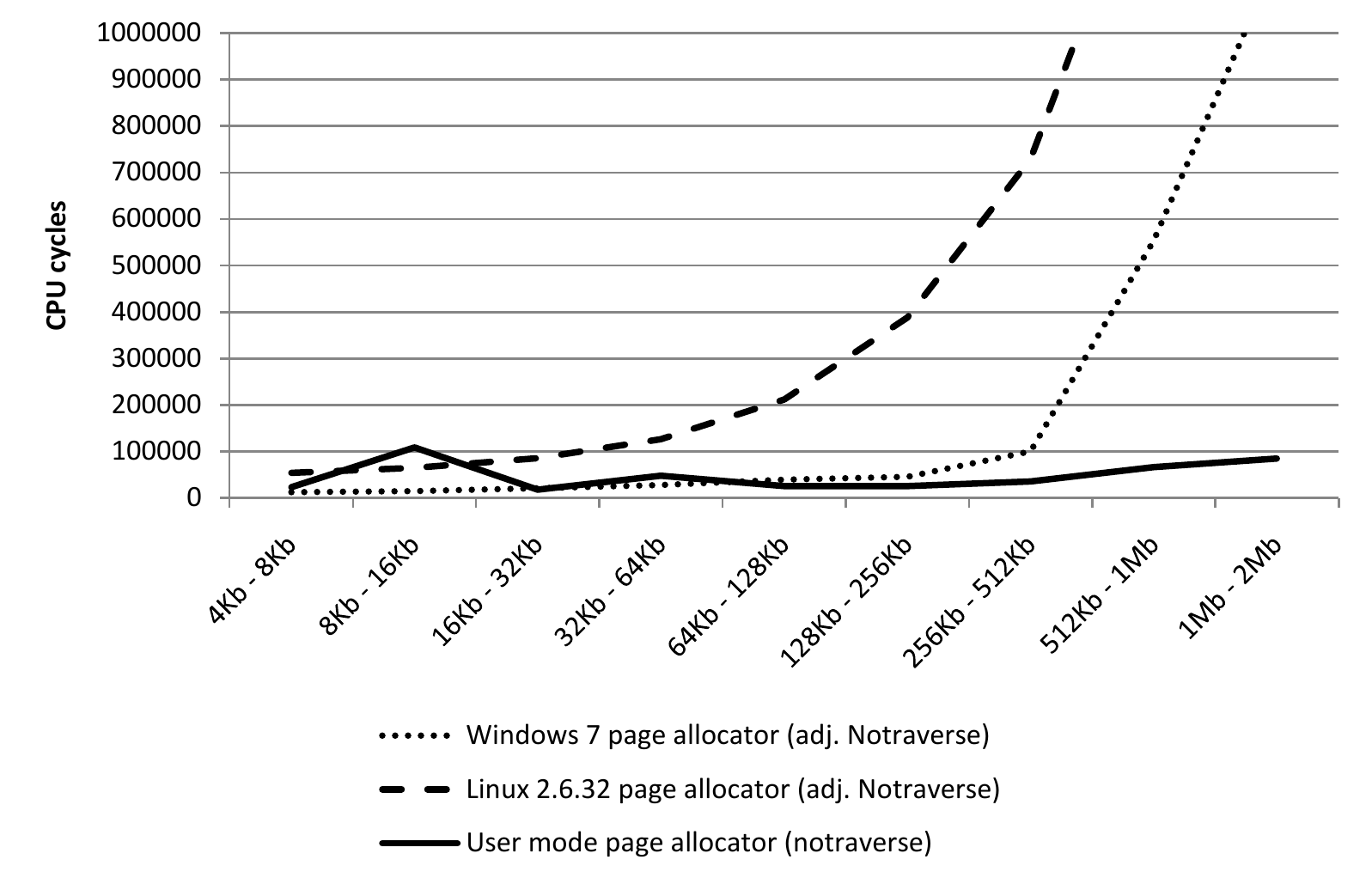}
  \caption{An adjusted for traversal costs log-linear plot of how the system and user mode page allocators scale to block sizes between 4Kb and 1Mb.}
  \label{FigSystemVsUserModeBreakout}
\end{figure}

Breaking out the range 4Kb--1Mb on a log-linear plot in Figure \ref{FigSystemVsUserModeBreakout}, one can see very clear O(N) complexity for the Windows and Linux kernel page allocators and an approximate O(1) complexity for the user mode page allocator. There is no good reason why O(1) complexity should break down after 1Mb except for the highly inefficient method the test uses to access the MMU hardware (i.e. how the AWE functions are used).

\subsection{Effects of user mode page allocation on general purpose application allocators}

\begin{figure}[h]
  \centering
    \includegraphics[width=0.5\textwidth]{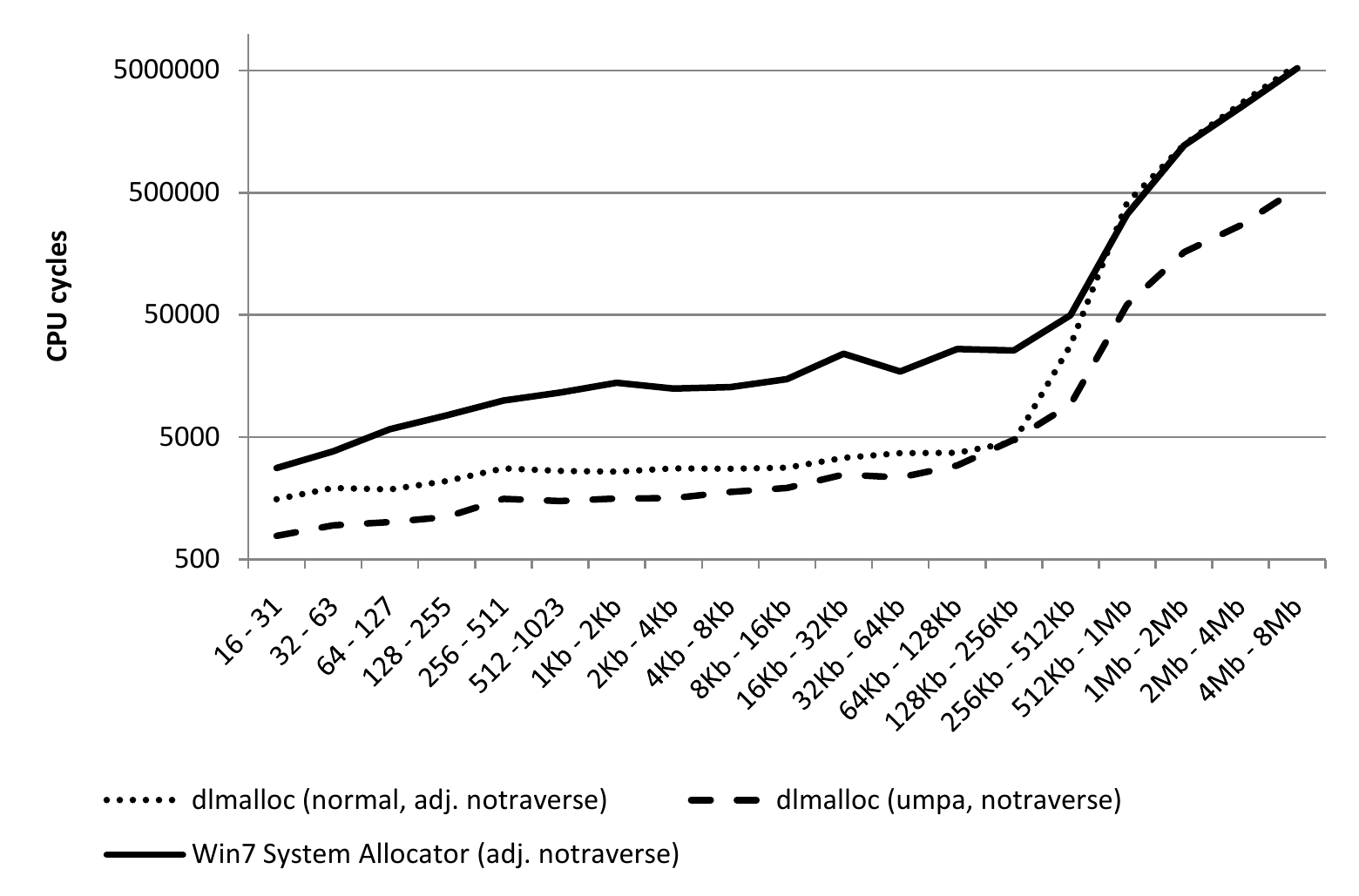}
  \caption{An adjusted for traversal costs log-log plot of how general purpose allocators scale to block sizes between 4Kb and 8Mb when running either under the system page allocator or the user mode page allocator (umpa).}
  \label{FigDlmallocSystemVsUserMode}
\end{figure}

Figure \ref{FigDlmallocSystemVsUserMode} shows how dlmalloc performs when running under the Windows kernel page allocator and the user mode page allocator with the performance of the Windows system allocator added for informational purposes. dlmalloc, under default build options, uses \texttt{mmap()}/\texttt{VirtualAlloc()} when the requested block size exceeds 256Kb and the curve for dlmalloc under the kernel page allocator clearly joins the same curve for the system allocator at 256Kb. Meanwhile, the user mode page allocator based dlmalloc does somewhat better.

\begin{figure}[h]
  \centering
    \includegraphics[width=0.5\textwidth]{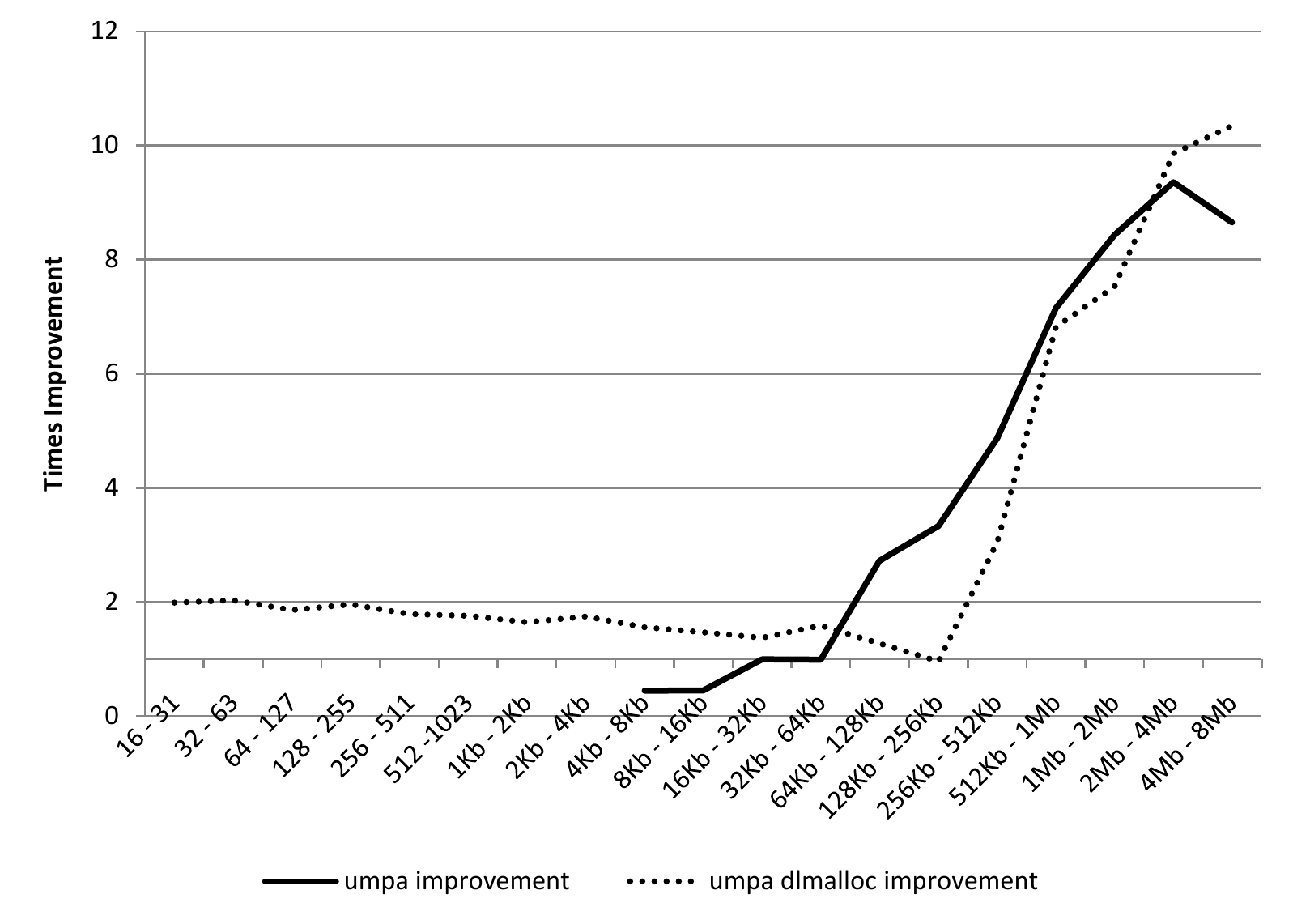}
  \caption{A summary of the performance improvements in allocation and frees provided by the user mode page allocator (umpa) where 1.0 equals the performance of the system page allocator.}
  \label{FigTimesImprovement}
\end{figure}

Breaking out this performance onto a linear scale where 1.0 equals the performance of the Windows kernel page allocator, Figure \ref{FigTimesImprovement} illuminates a surprising 2x performance gain for very small allocations when running under the user mode page allocator with a slow decline in improvement as one approaches the 128Kb--256Kb range. This is particularly surprising given that the test randomly mixes up very small allocations with very big ones, so why there should be a LOG(allocation size) related speed-up is strange. If one were to speculate on the cause of this, it would be suggested that the lack of cache pollution introduced by the lack of a page fault handler being called for every previously unaccessed page is most likely to be the cause.

\subsection{Effects of user mode page allocation on block resizing}

\begin{figure}[h]
  \centering
    \includegraphics[width=0.5\textwidth]{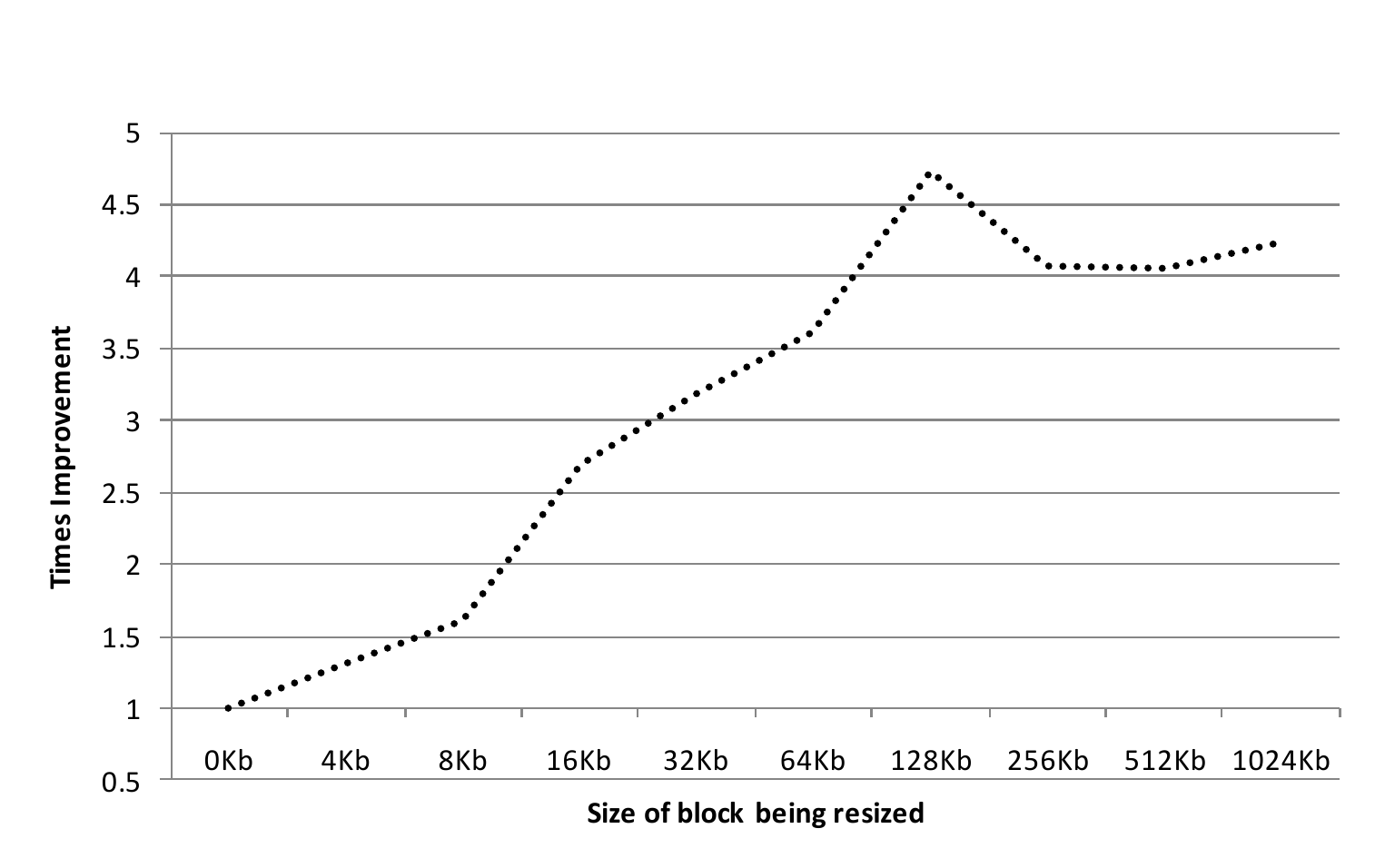}
  \caption{A summary of the performance improvements in block resizing provided by the page remapping facilities of the user mode page allocator where 1.0 equals the performance of (allocate new block + copy in old block + free old block).}
  \label{FigReallocTimesImprovement}
\end{figure}

Figure \ref{FigReallocTimesImprovement} shows no surprises considering the earlier results. The ability to remap pages and therefore avoid memory copies allows scale invariant \texttt{realloc()} performance up to 128Kb block sizes after which the prototype user mode page allocator joins the same scaling curve as traditional memory copy based \texttt{realloc()}. The implementation of \texttt{userpage\_realloc()} used in the prototype is extremely inefficient -- considerably worse than the implementations of \texttt{userpage\_malloc()} and \texttt{userpage\_free()} -- having to perform some four AWE function calls per item remapped. There is absolutely no reason why, if given proper operating system support, this operation would not continue to be scale invariant up to one quarter the level of that for allocations and frees rather than the $\left.^1\middle/ _{20}\right.$th seen here.

\subsection{Effects of user mode page allocation on real world application performance}

Using the equation from Gene Amdahl's 1967 paper \cite{amdahl1967validity} where the overall speedup from an improvement is $\frac{1}{(1-P)+\frac{P}{S}}$ where $P$ is the amount of time in the allocator and $S$ is the speedup of the allocator, if an application spends 5\% of its time in the memory allocator then doubling allocator performance will only yield at most a 2.56\% improvement. In order to determine the effects of user mode page allocation on present and future real world application usage profiles, a binary patcher was used to replace calls to the C allocator API as well as \texttt{VirtualAlloc()} et al. with calls to dlmalloc and the user mode page allocator respectively in a set of otherwise unmodified applications.

One must therefore bear in mind that the following results are therefore necessarily \textbf{a worst case scenario} for the user mode page allocator as it is used identically to the kernel page allocator. One should also consider that the Microsoft Windows kernel page allocator has a granularity of 64Kb, and that therefore the user mode page allocator is about the same speed at that granularity as shown in Figure \ref{FigSystemVsUserMode}.

The following tests were conducted with the results shown in Table \ref{TableRealWorldTestResults}:
\begin{itemize}
\item Tests 1 and 2: The x86 MinGW build of the GNU C++ compiler v4.5 was used to compile two projects which use the Boost Spirit Classic C++ recursive descent parser generator library v1.43. These tests were chosen as good examples of modern C++ code using extensive template meta-programming, and were conducted by compiling a file which uses the C preprocessor to include all the source files of the project into a single compiland. The first test compiles an ISO C grammar parser \cite{kaiser2004} which is a relatively easy test typical of current workloads, whereas the second test compiles most of an ISO C++98 grammar parser \cite{scalpel2010} written in very heavy C++0x metaprogramming. This second test is likely indicative of compiler workloads in a few years time.

Both tests were conducted with variation in flags, with (a) `-O0' (no optimisation), (b) `-g -O0' (debug info, no optimisation), (c) `-O3' (full optimisation) and (d) `-g -O3' (debug info, full optimisation) chosen. As one can see from the results in Table \ref{TableRealWorldTestResults}, the improvement given by the user mode page allocator appears to be much more dependent on compiler flags rather than memory usage with the least improvement given by turning on optimisation and the most improvement given by turning on the inclusion of debug information.

\item Test 3: A VBA automation script was written for Microsoft Word 2007 x86 which times how long it takes to load a document, paginate it ready for printing and publish the document as a print-ready PDF before closing the document. Document (a) is a 200,000 word, five hundred page book with a large number of embedded graphs and high resolution pictures, whereas document (b) is a set of twenty-one extremely high resolution forty megapixel (or better) photographs. Surprisingly, the user mode page allocator gave a net performance decrease for document (a) and barely affected the result for document (b), despite that in the case of document (b) that Word is definitely making twenty-one large allocations and frees.

\item Test 4: The standard x86 MSVC build of Python v2.6.5 was used to run two benchmarks taken from the Computer Language Benchmarks Game \cite{computerlanguagebenchmarks2010}. Two programs were benchmarked: (a) \texttt{binary-trees} which is a simplistic adaptation of Hans Boehm's GCBench where $N$ number of binary trees are allocated, manipulated and deallocated and (b) \texttt{k-nucleotide} which reads in $N$ nucleotides from a FASTA format text file into a hash table, and then to count and write out the frequency of certain sequences of DNA. For test (a) iterations were performed for $13<=N<=19$ and for test (b) iterations were performed for $1,000,000<=N<=10,000,000$.

\item Test 5: For the final test it was thought worthwhile to try modifying the source code of a large application to specifically make use of the O(1) features of the user mode page allocator. Having access to the source code of a commercial military modelling and simulation application \cite{allen2005implementation} which is known to make a series of expansions of large arrays, the core Array container class was replaced with an implementation using \texttt{realloc()} and three of its demonstration benchmarks performed: (a) Penetration Grid (b) Tank and (c) Urban Town.
\end{itemize}

\begin{table}[h]
\caption{Effects of the user mode page allocator on the performance of selected real world applications.}
\begin{center}
\begin{tabular}{ r | r r }
& Peak Memory & \\
Test & Usage & Improvement \\
\hline
Test 1a (G++): & 198Mb & +2.99\% \\
Test 1b (G++): & 217Mb & +1.19\% \\
Test 1c (G++): & 250Mb & +5.68\% \\
Test 1d (G++): & 320Mb & +4.44\% \\
Test 2a (G++): & 410Mb & +3.04\% \\
Test 2b (G++): & 405Mb & +1.20\% \\
Test 2c (G++): & 590Mb & +5.25\% \\
Test 2d (G++): & 623Mb & +3.98\% \\
Test 3a (Word): & 119Mb & \textbf{-4.05\%} \\
Test 3b (Word): & 108Mb & +0.67\% \\
& & [-0.25\% - +1.27\%], \\
Test 4a (Python): & 6 - 114Mb & avrg. +0.47\% \\
& & [-0.41\% - +1.73\%], \\
Test 4b (Python): & 92 - 870Mb & avrg. +0.35\% \\
Test 5a (Solver): & -- & +1.41\% \\
Test 5b (Solver): & -- & +1.07\% \\
Test 5c (Solver): & -- & +0.58\% \\
\hline
\multicolumn{3}{r}{\emph{Mean = +1.88\%}} \\
\multicolumn{3}{r}{\emph{Median = +1.20\%}} \\
\multicolumn{3}{r}{\emph{Chi-squared probability of independence $p$ = 1.0}} \\
\end{tabular}
\end{center}
\label{TableRealWorldTestResults}
\end{table}

\begin{figure}[h]
  \centering
    \includegraphics[width=0.5\textwidth]{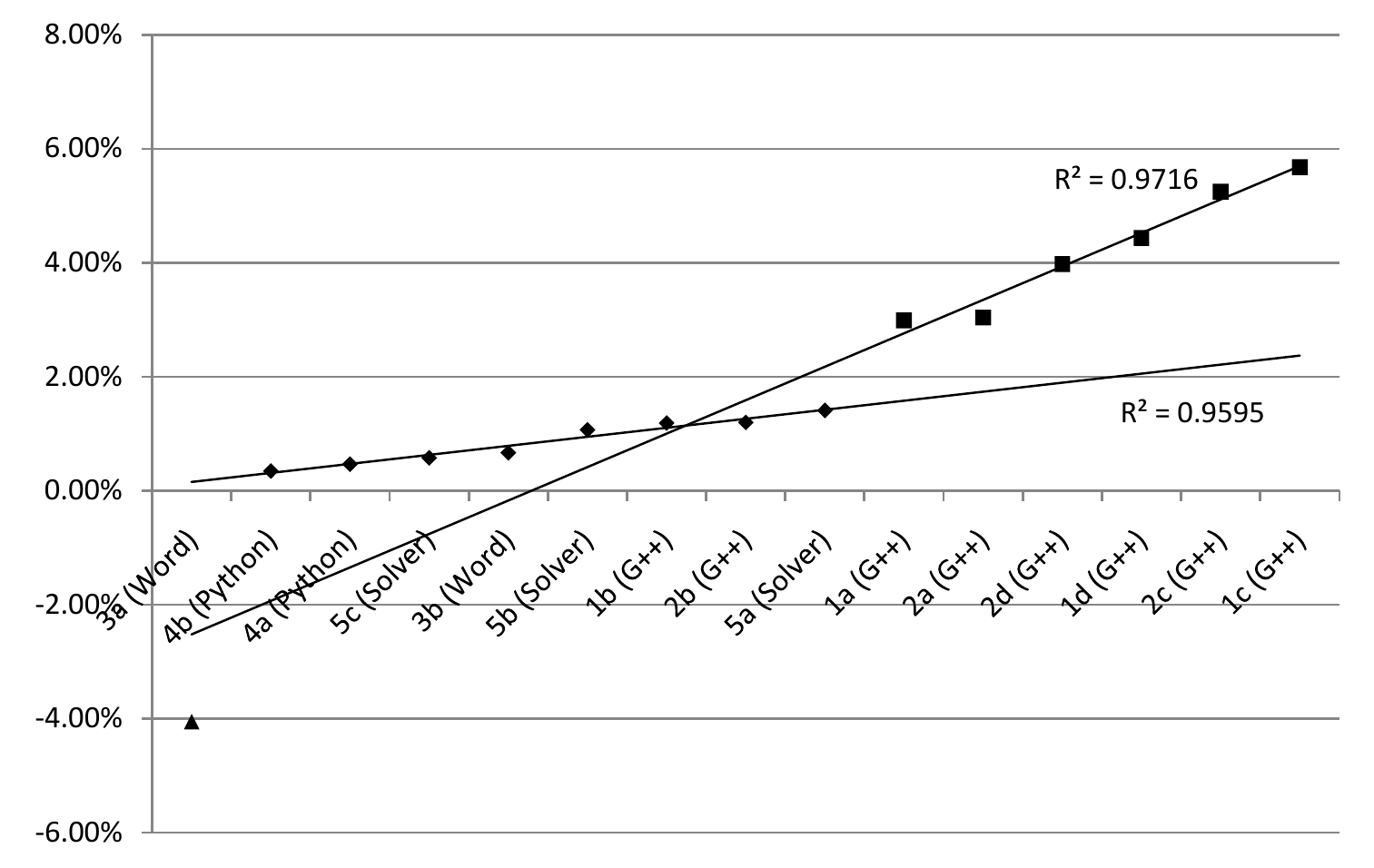}
  \caption{An illustration of structure present in the results shown by Table \ref{TableRealWorldTestResults}.}
  \label{FigRealWorldResultsStructure}
\end{figure}

Figure \ref{FigRealWorldResultsStructure} sorts these results and shows linear regression lines for the clusters subsequently identified. The high line -- entirely that of g++ -- displays a clear relationship, whereas the lower line is much more of a mix of applications. This structure, in addition to the total lack of correlation between the improvements and memory usage, strongly suggests that user mode page allocation has highly task-specific effects which are not general in nature. In other words, user mode page allocation affects a minority of operations a lot while having little to no effect on anything else.

As mentioned earlier, even half a percent improvement in a real world application can be significant depending on how that application uses its memory allocator. Apart from one of the tests involving Microsoft Word 2007, all of the tests showed a net positive gain when that application -- otherwise unmodified except for Test 5 -- is running under the prototype user mode page allocator. Given a properly implemented user mode page allocator with full operating system and batch API support, one could only expect these results to be significantly improved still further though still by no more than a few percent overall in the examples above.

\section{Conclusion}

To quote P.J. Denning (1996) \cite{denning1996} as he reflected on the history of computer memory management:
\begin{quote}
\emph{``From time to time over the past forty years, various people have argued that
virtual memory is not really necessary because advancing memory technology
would soon permit us to have all the random-access main memory we could
possibly want. Such predictions assume implicitly that the primary reason for
virtual memory is automatic storage allocation of a memory hierarchy. The
historical record reveals, to the contrary, that the driving force behind virtual
memory has always been \textbf{simplifying programs} (and programming) by
insulating algorithms from the parameters of the memory configuration and by
allowing separately constructed objects to be shared, reused, and protected. The
predictions that memory capacities would eventually be large enough to hold
everything have never come true and there is little reason to believe they ever
will. And even if they did, each new generation of users has discovered that its
ambitions for sharing objects led it to virtual memory. Virtual memory
accommodates essential patterns in the way people use computers. It will still be
used when we are all gone."} (pp. 13--14, emphasis added).
\end{quote}

It would be churlish to disagree with the great Prof. Denning, and it is hoped that no one has taken this paper to mean this. Rather, the time series data presented in Figures \ref{FigSSDsVsHardDrives}--\ref{FigMemorySizeVsSpeed} were intended to show that the rate of growth of growth (the second derivative) in RAM capacity has been outstripping the rate of growth of growth in magnetic storage since approximately 1997 -- this being a fact unavailable to Prof. Denning at the time he wrote the quote above. Such a fundamental change must both introduce new inconveniences for programmers and remove old ones as the effort-to-reward and cost-to-benefit curves shift dramatically. This paper has proposed the technique of user mode page allocation as one part of a solution to what will no doubt be a large puzzle.

Nothing proposed by this paper is a proposal against virtually \emph{addressed} memory: this is extremely useful because it lets programmers easily program today for tomorrow's hardware, and it presents an effectively uniform address space which is easy to conceptualise mentally and work with via pointer variables which are inherently one dimensional in nature. Back in the bad old days of segment based addressing, one had a series of small ranges of address space much the same as if one disables the MMU hardware and works directly with physical RAM today. Because the introduction of a MMU adds considerable latency to random memory access -- especially so when that MMU is hardware virtualised as this paper proposes -- one must always ask the question: \emph{does the new latency added here more than offset the latency removed overall?}

The central argument of this paper is that if a user mode page allocator is \textbf{already} no slower on otherwise unmodified existing systems today, then the considerable simplifications in implementation it allows across the entire system ought to both considerably improve performance and memory consumption at a geometrically increasing rate if technology keeps improving along historical trends. Not only this, but millions of programmer man hours ought to be saved going on into the future by avoiding the writing and debugging of code designed to reuse memory blocks rather than throwing them away and getting new ones. One gets to make use of the considerably superior metadata about memory utilisation available at the application layer. And finally, one opens the door wide open to easy and transparent addition of support for trans-cluster NUMA application deployments, or other such hardware driven enhancements of the future, without needing custom support in the local kernel.

This paper agrees entirely with the emphasised sentiment in Prof. Denning's quote: generally speaking, \emph{the simpler the better}. This must be contrasted against the reality that greater software implementation parallelization will be needed going on into the future, and parallelizability tends to contradict sharing except in read-only situations. Therefore, one would also add that generally speaking, \emph{the simpler and fewer the interdependencies the better} too because the more parallelizable the system, the better it will perform as technology moves forward. To that end, operating system kernels will surely transfer ever increasing amounts of their implementation into the process space to the extent that entering the kernel only happens when one process wishes to establish or disconnect communications with another process and security arbitration is required. Given the trends in hardware virtualisation, it is likely that simple hardware devices such as network cards will soon be driven directly by processes from user mode code and the sharing of the device will be implemented in hardware rather than in the kernel. One can envisage a time not too far away now where for many common application loads, the kernel is entered once only every few seconds -- an eternity in a multi-gigahertz world.

Effectively of course this is a slow transition into a `machine of machines' where even the humble desktop computer looks more like a distributed multiprocessing cluster rather than the enhanced single purpose calculator of old. We already can perceive the likely future of a thin hypervisor presiding over clusters of a few dozen specialised stream and serial processing cores, with each cluster having a few dozen gigabytes of local memory with non-unified access to the other clusters' memory taking a few hundred cycles longer. How can it be wise to consider the paradigm espoused by the 1970s C malloc API as even remotely suitable for such a world? Yet when you examine how memory is allocated today, it very rarely breaks the C memory allocation paradigm first codified by the Seventh Edition of Unix all the way back in 1979 \cite{seventhunix1979}.

Surely it has become time to replace this memory allocation paradigm with one which can handle non-local, hardware assisted, memory allocation as easily as any other? Let the application have direct but virtualised access to the bare metal hardware. Let user mode page allocation replace kernel mode page allocation!

\subsection{Future and subsequent work}
Without doubt, this paper raises far more questions than answers. While this paper has tried to investigate how user mode page allocation would work in practice, there is no substitute for proper implementation. Suitable syscall support could be easily added to the Linux or Windows kernels for MMU virtualisation capable hardware along with a kernel configuration flag enabling the allocation of physical memory pages to non-superusers. One would then have a perfect platform for future allocator research, including those barely imaginable today.

The real world application testing showed an up to 5\% performance improvement when running under the user mode page allocator. How much of this is due to lack of cache pollution introduced by the page fault handler, or the scale invariant behaviour of the user mode page allocator up to 1Mb block sizes, or simply due to differences in free space fragmentation or changes in cache locality is unknown. Further work ought to be performed to discover the sources of the observed performance increases, as well as determining whether the aggregate improvements are introduced smoothly across whole program operation or whether in fact some parts of the application run slower and are in aggregate outbalanced by other parts running quicker.

The present 1979 C malloc API has become a significant and mostly hidden bottleneck to application performance, particularly in object orientated languages because (i) object containers have no way of batching a known sequence of allocations or deallocations e.g. \texttt{std::list<T>(10000)} or even \texttt{std::list<T>::{\raise.17ex\hbox{$\scriptstyle\mathtt{\sim}$}}list()} (ii) \texttt{realloc()} is incapable of inhibiting address relocation which forces a copy or move construction of every element in an array when resized \cite{kimpe2006, buhr2010}. Considering how much more important this overhead would become under a user mode page allocator, after writing this paper during the summer of 2010 I coordinated the collaborative development of a ``latency reducing v2 malloc API" for C1X. After its successful conclusion, a proposal for a change to the ISO C1X library specification standard was constructed and submitted to WG14 as N1527 \cite{n1527} as well as two C99 reference implementations of N1527 were written and made available to the public \cite{c1x_n1527}. This proposal adds batch memory allocation and deallocation along with much finer hinting and control to the allocator and VM implementations. As an example of what N1527 can bring to C++, a simple \texttt{std::list<int>(4000000)} runs 100,000 times faster when \texttt{batch\_alloc1()} is used for construction instead of multiple calls to operator new! I would therefore urge readers to take the time to examine N1527 for deficiencies and to add support for its proposed API to their allocators.

\acks
The author would like to thank Craig Black, Kim J. Allen and Eric Clark from Applied Research Associates Inc. of Niceville, Florida, USA for their assistance during this research, and to Applied Research Associates Inc. for sponsoring the development of the user mode page allocator used in the research performed for this paper. I would also like to thank Doug Lea of the State University of New York at Oswego, USA; David Dice from Oracle Inc. of California, USA; and Peter Buhr of the University of Waterloo, Canada for their most helpful advice, detailed comments and patience.

\bibliographystyle{unsrtnat}
\softraggedright
\bibliography{MemoryAllocation,ThisPaper}

\end{document}